\documentclass[11pt,preprint]{aastex}

\usepackage{amsmath}

\shorttitle{H$_2$O-Ice Crystallinity on TNOs}

\shortauthors{Terai et al.}

\begin{document}

\title{Photometric Measurements of H$_2$O Ice Crystallinity on Trans-Neptunian Objects \altaffilmark{*}}

\altaffiltext{*}{Based on data collected at Subaru Telescope, which is operated by the National
Astronomical Observatory of Japan (NAOJ).
}

\author{Tsuyoshi Terai$^{1}$, Yoichi Itoh$^{2}$, Yumiko Oasa$^{3}$, Reiko Furusho$^{4}$,
and Junichi Watanabe$^{4}$}
\affil{
$^1$ Subaru Telescope, National Astronomical Observatory of Japan, 650 North A`ohoku Place, Hilo,
HI 96720, USA\\
$^2$ Center for Astronomy, University of Hyogo, 407-2 Nishigaichi, Sayo-cho, Sayo-gun, Hyogo
679-5313, Japan\\
$^3$ Faculty of Education, Saitama University, 255 Shimo-Okubo, Sakura, Saitama 338-8570, Japan\\
$^4$ National Astronomical Observatory of Japan, 2-21-1 Osawa, Mitaka, Tokyo 181-8588, Japan\\
}
\email{tsuyoshi.terai@nao.ac.jp}

\begin{abstract}

We present a measurement of H$_2$O ice crystallinity on the surface of trans-neptunian objects
(TNOs) with near-infrared narrow-band imaging.
The newly developed photometric technique allows us to efficiently determine the strength of
an 1.65-$\micron$ absorption feature in crystalline H$_2$O ice.
Our data for three large objects, Haumea, Quaoar, and Orcus, which are known to contain
crystalline H$_2$O ice on the surfaces, show a reasonable result with high fractions of the
crystalline phase.
It can also be pointed out that if the H$_2$O-ice grain size is larger than $\sim$20~$\micron$,
the crystallinities of these objects are obviously below 1.0, which suggest the presence of the
amorphous phase.
Especially, Orcus exhibits a high abundance of amorphous H$_2$O ice compared to Haumea and
Quaoar, possibly indicating a correlation between bulk density of the bodies and surface
crystallization degree.
We also found the presence of crystalline H$_2$O ice on Typhon and 2008~AP$_{129}$, both of
which are smaller than the minimum size limit for inducing cryovolcanism as well as a transition
from amorphous to crystalline through the thermal evolution due to the decay of long-lived
isotopes.

\end{abstract}


\section{INTRODUCTION}\label{sec01}

Trans-neptunian objects (TNOs) are believed to be ice-rich bodies formed in regions distant from
the Sun.
Several kinds of icy species have been detected from TNOs by previous visible and
near-infrared spectroscopic observations.
H$_2$O ice, known as the most abundant volatile material in the solar system, is generally the
primary component on icy surfaces of TNOs except for the largest objects such as Pluto, Eris, Sedna,
and Makemake, that are covered by CH$_4$ ice
\citep[e.g.,][]{1993Sci...261..745O, 2005ApJ...635L..97B, 2005A&A...439L...1B,2006A&A...445L..35L}.
H$_2$O ice is condensed from the vapor in nebula gas as amorphous phase in a cold environment
far bellow 100~K, \citep{1998ASSL..227..139J,2008Icar..197..307M}.
Transition from amorphous to crystalline ice requires $\sim$1~hr at 130~K, while $\sim$10$^3$~yr at
90~K \citep{1996ApJ...473.1104J}.
Pristine H$_2$O ice maintains an amorphous state from the beginning of the solar system if the
temperature remains lower than $\sim$80~K.
Although the typical surface temperature of TNOs is $\sim$40--60~K \citep{2008ssbn.book..161S}, the
spectra of large TNOs obviously exhibit the absorption feature of crystalline H$_2$O ice at
1.65~$\micron$, e.g., 
Charon \citep{2000Sci...287..107B,2000Icar..148..324B},
Quaoar \citep{2004Natur.432..731J,2009A&A...501..349D}, 
Haumea \citep{2006ApJ...640L..87B,2007ApJ...655.1172T,2007A&A...466.1185M}, and
Orcus \citep{2005A&A...437.1115D,2008A&A...479L..13B}.
Haumea's satellite Hi`iaka \citep{2011A&A...528A.105D} and collisional family members
\citep{2006ApJ...640L..87B} are also known to be covered by crystalline H$_2$O ice.
Additionally, several moderate-size TNOs/Centaurs have been suggested to have surfaces covered in
crystalline H$_2$O ice \citep[e.g.,][]{2011Icar..214..297B}.

These objects are highly likely to have experienced a certain process resulting in the production
and/or provision of crystalline H$_2$O ice on their surfaces.
Several mechanisms have been suggested, including radiogenic heating
\citep{2006Icar..183..283M,2011A&A...529A..71G}, cryovolcanism
\citep{2007ApJ...663.1406C,2009Icar..202..694D}, and micrometeorite impact annealing
\citep{2010Icar..208..492P}.
However, it remains unknown which of them primarily causes the presence of crystalline H$_2$O ice
on the surfaces.
Elucidation of the generation process could provide an insight into the formation, evolution, and
possibly interior structure of icy small bodies in the outer solar system.
For this purpose, it is essential to clarify the following ambiguous questions; 
(i) Is the presence of crystalline H$_2$O ice on the surfaces ubiquitous among TNOs?
If yes, is the crystallinity constant among them?
(i\hspace{-0.5pt}i) Does the presence/abundance of crystalline ice depend on the body size and/or
other parameters?
In the former case, what is the size limit for the surface containing crystalline ice?

In the present circumstances, however, the general knowledge regarding abundance distribution of
surface crystallinity in the TNO population is still poor because of a lack of observational
constraints.
Most of the known TNOs are too faint to obtain spectra with sufficient quality to
accurately determine the ratio of crystalline and amorphous phases on their surfaces even by
largest-class telescopes.

In this paper, we introduce a new measurement technique for H$_2$O ice crystallinity on icy small
bodies with near-infrared photometric data acquired by the Subaru telescope.
We use the narrow-band filter called ``NB1657", allowing us to evaluate the strength of
1.65-$\micron$ absorption of crystalline H$_2$O with high precision.
The results demonstrate that this method is useful for an effective survey of the crystalline ice
abundance from a large number of TNOs.

\section{OBSERVATIONS AND DATA REDUCTION}\label{sec02}

\begin{figure}
\epsscale{0.90}
\plotone{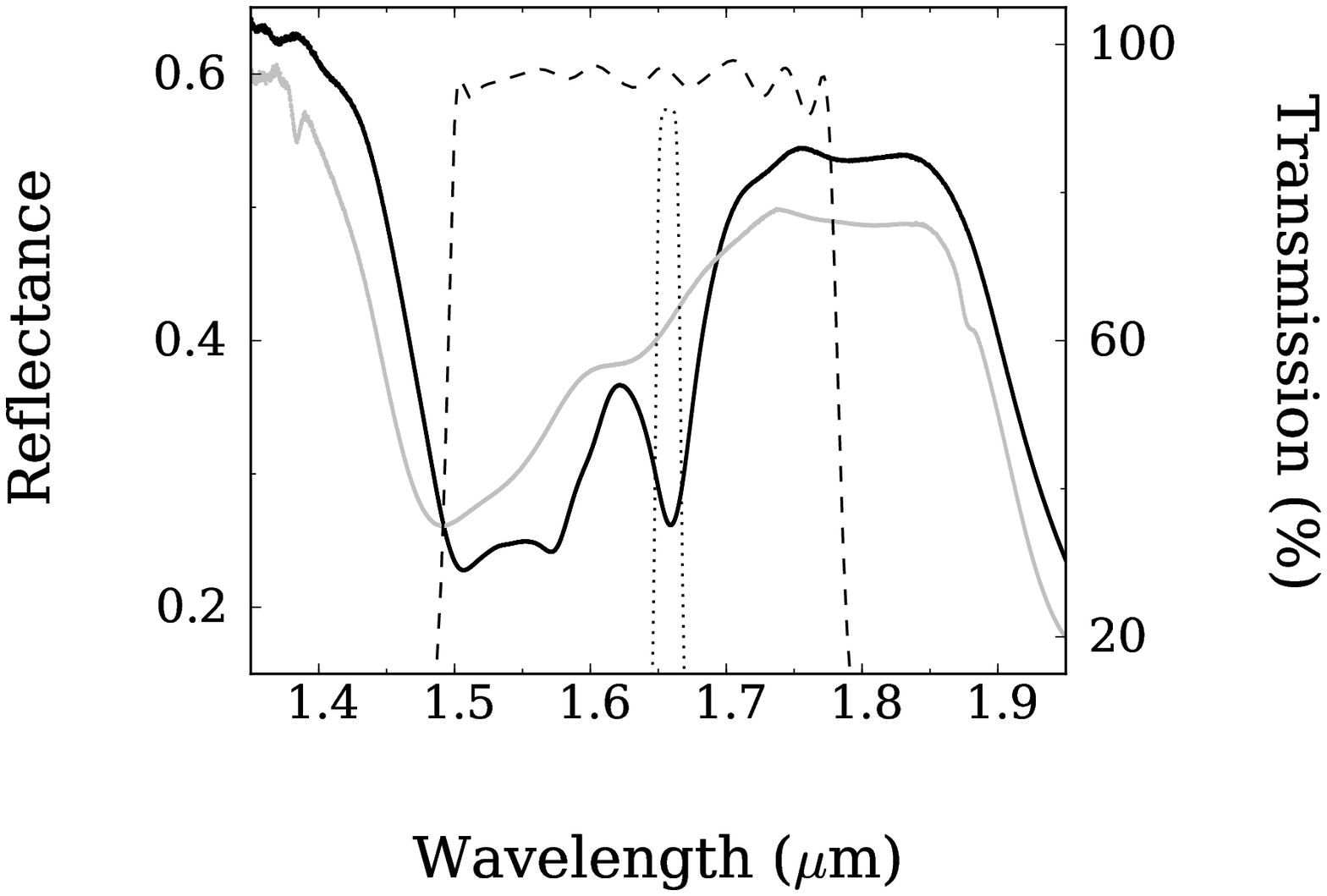}
\caption{
 Near-infrared reflectance spectra of H$_2$O ice in crystalline (solid line) and amorphous
 (gray line) states \citep[from][]{2008Icar..197..307M}.
 The dashed and dotted curves show transmission curves of the $H$-band and NB1657 filters
 installed in MOIRCS, respectively
 \label{fig01}}
\end{figure}

Our observation was carried out on April 7, 2013, with Multi-Object InfraRed Camera and
Spectrograph \citep[MOIRCS;][]{2006SPIE.6269E..38I,2008PASJ...60.1347S} mounted on the 8.2-m
Subaru telescope.
MOIRCS consists of two 2048$\times$2048 arrays with pixel scale of $0\farcs117$, each of which
covers a $4\arcmin \times 3\farcm5$ field of view.
We performed a near-infrared imaging using the $H$ band and narrow band filter NB1657.
The NB1657 with the center wavelength of 1.657~$\micron$ and band width of 0.019~$\micron$
\citep{2014ApJ...789...18K} is suitable for diagnosing the absorption of crystalline H$_2$O ice
at 1.65~$\micron$ (see Figure~\ref{fig01}).
The sky condition was photometric, with a seeing size of mostly 0\farcs6--0\farcs8.

Our target objects consist of seven TNOs listed in Table~\ref{tab1}.
These objects have been reported to contain H$_2$O ice on the surfaces in the near-infrared
spectroscopy by \citet{2012AJ....143..146B}, i.e., $f_{\rm H_2O}$~$\gtrsim$~0.1, where
$f_{\rm H_2O}$ represents the spectral fraction of H$_2$O ice.
The observational circumstances are shown in Table~\ref{tab2}.
Every image was taken with an exposure of 42--80~sec in the $H$ band and 90--180~sec in
the NB1657 under sidereal tracking.
The sky motion within the exposure time is sufficiently smaller than the seeing size to treat the
target objects as a point source.
The NB1657 data was obtained between just before/after $H$-band imagings to reduce color
uncertainty caused by rotational brightness variation.

Data reduction was conducted using IRAF produced by the National Optical Astronomy Observatories
(NOAO) and MCSRED\footnote{\url{http://www.naoj.org/staff/ichi/MCSRED/mcsred.html}}
\citep{2011PASJ...63S.415T} with standard processes: linearity correction, flat-fielding,
sky subtraction, and distortion correction.
Images with each band were shifted according to the sky motion of a target object and
were combined.
Position-matched composite images were also created for determining the point spread function
(PSF) from the field stars.
We adopted aperture photometry for flux measurement using the IRAF/APPHOT package.
To increase signal-to-noise ratio, the aperture correction technique has been applied.
Summed pixel counts within a small aperture with radius of $\sim$$0.5$--$0.8$~arcsec were
converted into total flux based on the flux ratio computed from the PSF profile given by the
field stars.
Haumea is an exception because of a lack of field stars and thus its flux was directly measured
with a large aperture (2.2~arcsec in radius).
The $H$-band magnitude and $H$$-$NB1657 index were calibrated with a G2V star,
2MASS~J06430376$-$0117471 ($H$ = 13.630 $\pm$ 0.031 mag) or 2MASS~J16375427+4052592
($H$ = 13.161 $\pm$ 0.027 mag), taken within 2.5~hours of target data acquisition.
$\Delta$($H$$-$NB1657) is defined as the offset $H$$-$NB1657 index that the Sun color is subtracted.

The results of photometry are shown in Table~\ref{tab3}.
Most of the $H$$-$NB1657 indexes have been determined with uncertainty of $\sim$0.02~mag or less,
allowing us to estimate abundance of crystalline H$_2$O ice on the surface of the target
objects.
Unfortunately, however, Ceto contains a large photometric error because of the insufficient
exposure time (500~sec with the $H$ and 720~sec with the NB1657 filters) due to restriction of the
observation time.
We exclude Ceto from the sample for the following analysis.

\section{RESULTS}\label{sec03}

\subsection{Model Spectra}\label{sec:subsec3-1}

From the measured $H$$-$NB1657 index, we obtain a fraction of the crystalline phase in the
H$_2$O ice spectrum, hereinafter called the ``crystallinity factor" ($f_{\rm crys}$).
In this paper, the reflectance spectrum of phase-mixed H$_2$O ice, $S_{\rm H_2O} (\lambda)$
($\lambda$ is wavelength in $\micron$), is represented by a linear sum of the crystalline and
amorphous spectra ($S_{\rm crys} (\lambda)$ and $S_{\rm amor} (\lambda)$, respectively) as
defined in \cite{2008Icar..193..397N}, i.e.,
\begin{equation} 
S_{\rm H_2O} (\lambda) = f_{\rm crys} \, S_{\rm crys} (\lambda)
                       + (1 - f_{\rm crys}) \, S_{\rm amor} (\lambda).
\label{eq1}
\end{equation}
Modeling reflectance spectra of the target objects is required to convert the 1.65-$\micron$
absorption strength derived from $\Delta$($H$$-$NB1657) into the crystallinity factor.

We assume that near-infrared spectra of those bodies are represented by a simple model
consisting of H$_2$O ice spectrum and linear continuum presented by \cite{2012AJ....143..146B}.
The model spectrum, $S (\lambda)$, is given as
\begin{equation} 
S (\lambda) = f_{\rm H_2O} \, S_{\rm H_2O} (\lambda)
+ (1 - f_{\rm H_2O}) \left[ m_{\rm cont} (\lambda - 1.74 \, \micron)
+ 0.49 \right],
\label{eq2}
\end{equation}
where $f_{\rm H_2O}$ is a spectral fraction of H$_2$O ice and $m_{\rm cont}$ is a continuum slope.
The H$_2$O ice spectra were generated from the geometric albedo $A_p$ described by the radiative
transfer model of \cite{1993tres.book.....H} as
\begin{equation} 
A_p \simeq r_0 \left( \frac{1}{2} + \frac{1}{6} r_0 \right)
    + \frac{w}{8} \left[ (1 + B_0) p(0) - 1\right],
\label{eq3}
\end{equation}
where $w$ is the single scattering albedo given from the optical constants and grain size,
$r_0$ is the diffusive reflectance given by $\frac{1 - \sqrt{1-w}}{1 + \sqrt{1-w}}$,
$B_0$ is the total amplitude of the opposition surge, and $p$(0) is the phase function at zero
phase angle.
We used $B_0$~=~0.67, the typical value among icy satellites \citep{1998ASSL..227..157V}
as in \cite{2009AJ....137..315M}, and isotopic scattering, i.e., $p$(0)~=~1.0.
The optical constants of amorphous and crystalline H$_2$O ices were derived from the laboratory
data provided by
\cite{2008Icar..197..307M}\footnote{Mastrapa, R. M. E., Optical Constants and Lab Spectra
of Water Ice V1.1. EAR-X-I1083-5-ICESPEC-V1.1. NASA Planetary Data System, 2012
(\url{http://sbn.psi.edu/pds/resource/icespec.html}).}.

The grain size is one of the most sensitive parameters to determine the reflectance spectrum
but still unknown.
We tentatively used uniform grain particles with a diameter of $d$~=~50~$\micron$ as in
\cite{2012AJ....143..146B}.
The uncertainty due to the dependency on grain size is discussed in Section~\ref{sec:subsec3-3}.

\begin{figure}
\epsscale{0.90}
\plotone{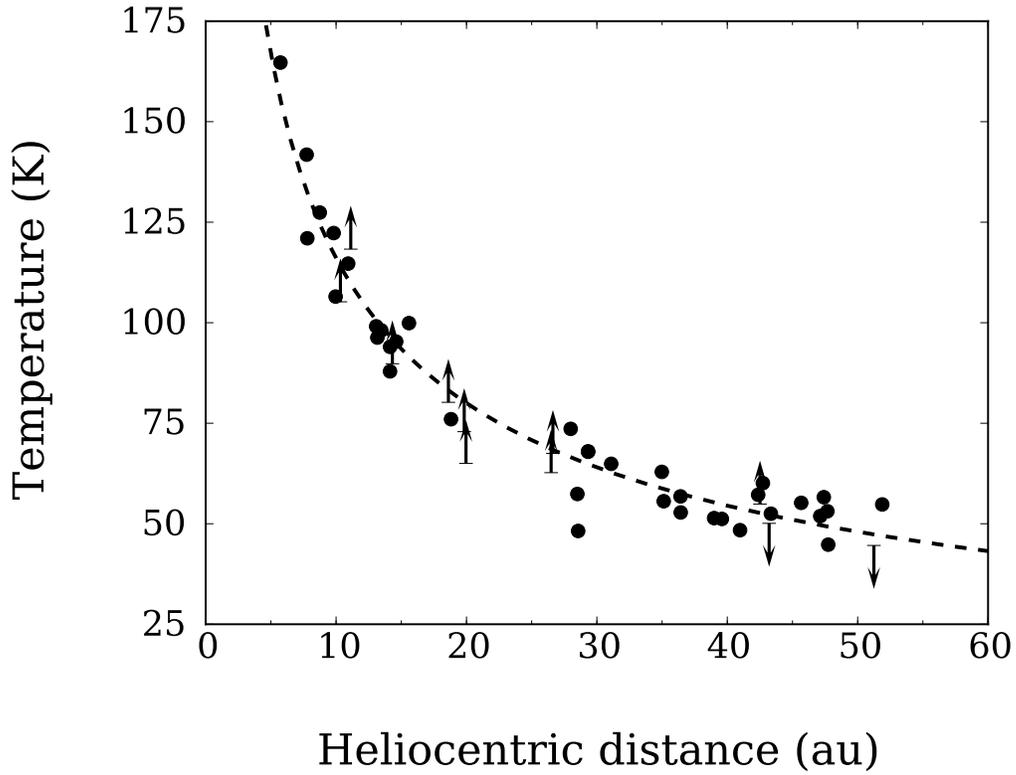}
\caption{
 Temperatures ($T$) of TNOs and Centaurs based on Spitzer Space Telescope 24- and 70-$\micron$ data
 \citep{2008ssbn.book..161S} as a function of the heliocentric distance ($r_{\rm h}$).
 The dashed line shows a best-fit expression of $T \propto r_{\rm h}^{-1/2}$ (see text).
 \label{fig02}}
\end{figure}

The absorption spectrum of H$_2$O ice, especially the 1.56-$\micron$ and 1.65-$\micron$ bands in
crystalline H$_2$O ice, varies with temperature \citep[e.g.,][]{1998JGR...10325809G}.
Surface temperature of airless solid bodies depends on the solar flux, rotation, and surface
properties.
\cite{2008ssbn.book..161S} presented temperatures of 49 TNOs/Centaurs derived from 24-$\micron$
and 70-$\micron$ flux data collected by the Spitzer Space Telescope.
Figure~\ref{fig02} shows the color temperatures as a function of target distance from the Sun
at the observations.
The plot is well approximated by an equation of
\begin{equation} 
T = (389 \pm 14) r_{\rm h}^{-1/2} + (-7.1 \pm 3.3),
\label{eq4}
\end{equation}
where $T$ and $r_{\rm h}$ are the temperature in kelvin and heliocentric distance in au,
respectively.
Here, we consider the thermal temperatures derived from this equation assuming isothermal
blackbodies as their ice temperatures.
Its appropriateness is assessed below.
Mastrapa's dataset contains the optical constants of crystalline H$_2$O ice at temperatures from
20--150~K every 10~K and those of amorphous H$_2$O ice at temperatures higher/lower than 70~K.
Based on the estimated temperatures, we drew the spectral reflectance of crystalline/amorphous
ices from the optimum optical constant data for each target object as seen in
Table~\ref{tab4}.
The model spectra of the target objects were generated from Equation~(\ref{eq2}) with the synthetic
spectra of H$_2$O ice as mixtures of crystalline and amorphous phases given by Equation~(\ref{eq1}).

The spectral fraction of H$_2$O ice $f_{\rm H_2O}$ and continuum slope $m_{\rm cont}$ were assumed to
be those suggested by \cite{2012AJ....143..146B} as listed in Table~\ref{tab1}.
Note that these parameters have been estimated from simple modeling with the optical constants of
fully crystallized ice.
We evaluated the variability of $f_{\rm H_2O}$ determined through the model depending on
crystallinity.
The synthetic spectra from Equation~(\ref{eq2}) were compared between pure crystalline ice and
amorphous-dominated/mixed ones by least-squares optimization under the same conditions as
\cite{2012AJ....143..146B}, i.e., temperatures of 50~K, grain size of 50~$\micron$, and
wavelength ranges of 1.45--1.80~$\micron$ and 1.95--2.30~$\micron$.
Table~\ref{tab5} shows the best-fit $f_{\rm H_2O}$ values for the spectra created with
$f_{\rm crys}$ of 0.00, 0.25, and 0.50.
Since there are only small differences from the given values in any $f_{\rm crys}$ cases, one can
see that $f_{\rm H_2O}$ has little dependency on crystallinity.
It indicates the validity of our crystallinity measurement based on the published $f_{\rm H_2O}$
values.

To compare with the photometric data, the model spectra were converted into $\Delta$($H$$-$NB1657)
by
\begin{equation} 
\Delta(H - {\rm NB1657}) = 2.5 \log \left[
\frac{\int R_{\rm NB}(\lambda) S(\lambda) \lambda F_{\lambda, \Sun}(\lambda) d\lambda
\ / \int R_H(\lambda) S(\lambda) \lambda F_{\lambda, \Sun}(\lambda) d\lambda}
{\int R_{\rm NB}(\lambda) \lambda F_{\lambda, \Sun}(\lambda) d\lambda
\ / \int R_H(\lambda) \lambda F_{\lambda, \Sun}(\lambda) d\lambda} \right],
\label{eq5}
\end{equation}
where $R_H(\lambda)$ and $R_{\rm NB}(\lambda)$ are the response functions for the $H$ and NB1657
bands, respectively.
$F_{\lambda, \Sun}(\lambda)$ is the wavelength flux density of the Sun.
The response functions include atmospheric transmission at the summit of Maunakea generated by
ATRAN modelling software \citep{Load92} assuming airmass of 1.0 and water vapor column of 1.0~mm.
We used the solar reference spectrum distributed by STScI Calibration Database
System\footnote{\url{http://www.stsci.edu/hst/observatory/crds/calspec.html}}.

\subsection{Crystallinity}\label{sec:subsec3-2}

\begin{figure}
\epsscale{0.90}
\plotone{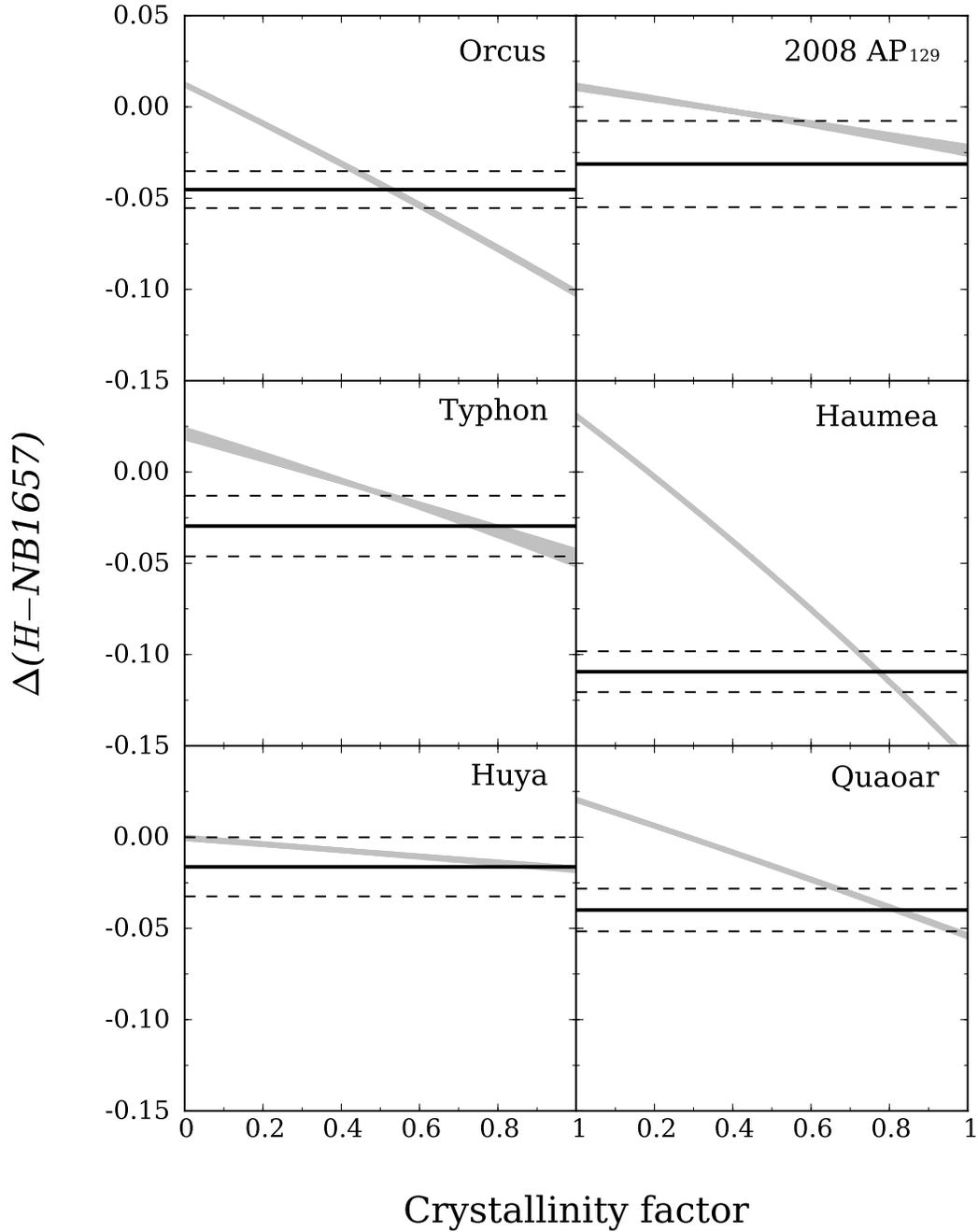}
\caption{
 Relative $H$$-$NB1657 indexes of the observed objects (solid lines) and 1~$\sigma$ errors of
 photometry (dashed lines).
 The gray lines show the relation between $H$$-$NB1657 index and crystallinity factor estimated
 from the model spectra.
 The width of gray lines represents uncertainty of the model parameters (see Table~\ref{tab1}). 
 \label{fig03}}
\end{figure}

By comparing the obtained $\Delta$($H$$-$NB1657) index with the model spectra, we determined the
crystallinity factors of H$_2$O ice for each target object.
Figure~\ref{fig03} shows $\Delta$($H$$-$NB1657) derived from our observation and modeling with
respect to the crystallinity factor.
The point where the model curve intersects with the measured value represents a plausible
crystallinity factor.
The resulting crystallinity factors are listed in Table~\ref{tab4}.
The determination accuracy depends on photometric precision and the slope of the model curve
given by abundance of H$_2$O ice.
For bright objects including Orcus, Haumea, and Quaoar, the crystallinity factor has been fixed
with accuracy of $\sim$0.1.
Huya is also bright, however, the error is very large due to low $f_{\rm H_2O}$
(0.08~$\pm$~0.02) causing a little $\Delta$($H$$-$NB1657) variation with crystallinity factor.
This implies that $f_{\rm H_2O} \sim 0.1$ is the limit of application of this technique
for TNOs in Subaru/MOIRCS observation.

Under hypothetical ice conditions with temperature based on thermal flux and grain size of
50~$\micron$, the model matching indicates abundant crystalline H$_2$O ice on the surfaces of
Haumea and Quaoar, as well as a moderate amount of the crystalline phase on Orcus.
This result agrees with the previous spectroscopic works showing a significant feature of the
1.65~$\micron$ absorption on those objects
\citep[e.g.,][]{2005A&A...437.1115D,2007ApJ...655.1172T,2004Natur.432..731J}.
Typhon also shows high crystallinity, consistent with the spectral model presented by
\cite{2009Icar..201..272G} (5\% of crystalline and 0\% of amorphous).

\begin{figure}
\epsscale{0.90}
\plotone{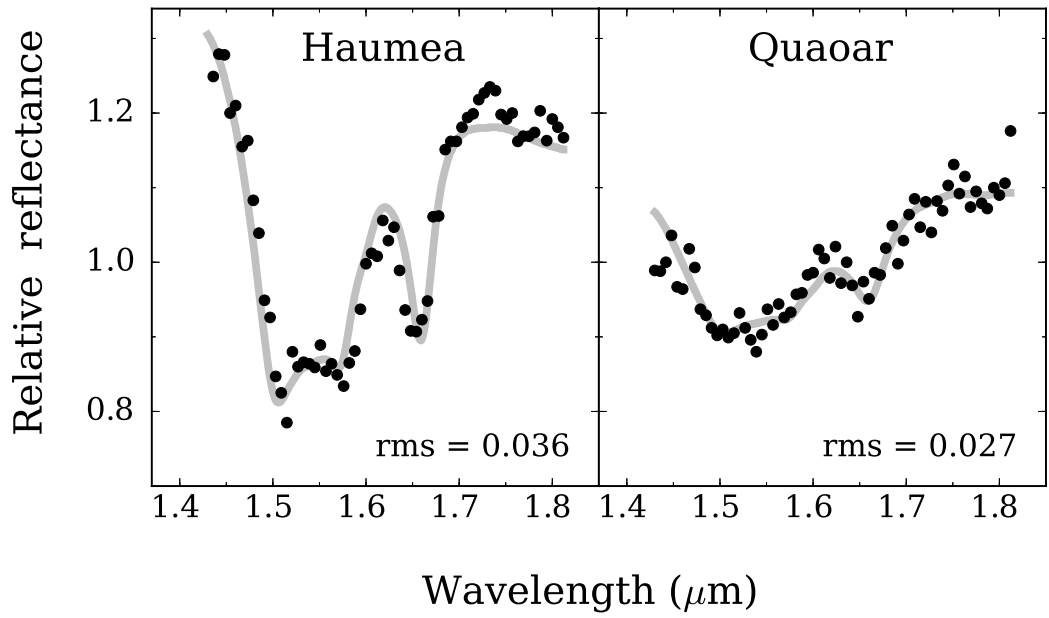}
\caption{
 Model spectra of Haumea (left) and Quaoar (right) with the determined fractions of crystalline
 H$_2$O ice (gray lines).
 The dots show the published spectral data for comparison.
 \label{fig04}}
\end{figure}

We evaluated the validity of this technique through comparison with published near-infrared spectra
of Haumea and Quaoar obtained with
Keck/NIRC\footnote{\url{http://web.gps.caltech.edu/$^\sim$pa/data/kbo\_info.html}}
\citep{2008AJ....135...55B}.
Figure~\ref{fig04} shows those spectra as well as the spectral models generated from
Equation~(\ref{eq2}) with the crystallinities measured by this work.
The models well reproduce the observed spectra inclusive of the 1.65-$\micron$ feature in both
objects.
It supports that our photometric method is useful to constrain the phase ratio of crystalline to
amorphous for an icy small body of which $f_{\rm H_2O}$ is known.
The effect of the assumed ice properties, temperature and grain size, for determination of the
crystallinity factor is examined in the following section.

\subsection{Ice Temperature And Grain Size}\label{sec:subsec3-3}

The depth and center wavelength of the 1.65-$\micron$ band sensitively vary with temperature and
grain size \citep{1975Icar...24..411F,1981JGR....86.3087C,1998JGR...10325809G,2012P&SS...61..124T},
while the two parameters were given as thermal temperature estimated from Equation~(\ref{eq4}) and
$d$~=~50~$\micron$, respectively, in the above modeling.
\cite{1999Icar..142..536G} presented disk-averaged H$_2$O ice temperatures of Jovian, Saturnian,
and Uranian satellites derived from their near-infrared spectra.
They pointed out that the ice temperature is generally lower than the brightness temperature
derived from thermal emission which is sensitive to warm regions on the surface with the typical
difference of $\sim$10~K.
The target objects could actually have a lower surface temperature.

We conducted additional modeling with the same processes presented in Section~\ref{sec:subsec3-1}
but using the optical constant data of crystalline H$_2$O ice at temperatures 10~K lower than those
in Table~\ref{tab4}.
The results, shown in Table~\ref{tab6}, indicate no significant change in the obtained
crystallinity factor compared with the original spectrum model in any objects.
Such a difference in the given temperature causes a systematic error of no more than $\sim$10\%
for our crystallinity measurements.

On the other hand, grain size can induce considerable uncertainty in the modeling.
Although the dominant size of the ice grains on surface layer of TNOs is still unknown, 
\cite{2011Icar..214..297B} reported surface composition models of 12 TNOs/Centaurs based on
their near-infrared spectroscopy data showing that most of their surfaces contain H$_2$O ice (in
crystalline and/or amorphous states) with the particle diameter of $\sim$10--200~$\micron$.
Many of the previous studies of TNO/Centaur spectra showed the best-matched grain sizes in
this range.
Note that H$_2$O ice with larger ice grains enhance the absorption coefficients, but the
enhancement saturates at the diameter of $\sim$1000~$\micron$.
In contrast, if the grains are smaller than $\sim$10~$\micron$, the absorptions are too weak to
measure the 1.65-$\micron$ feature via $\Delta$($H$$-$NB1657) index.
Thus, our technique is applicable to objects whose surfaces are dominated by H$_2$O ice with the
grain size of $\sim$10--1000~$\micron$.

\begin{figure}
\epsscale{0.90}
\plotone{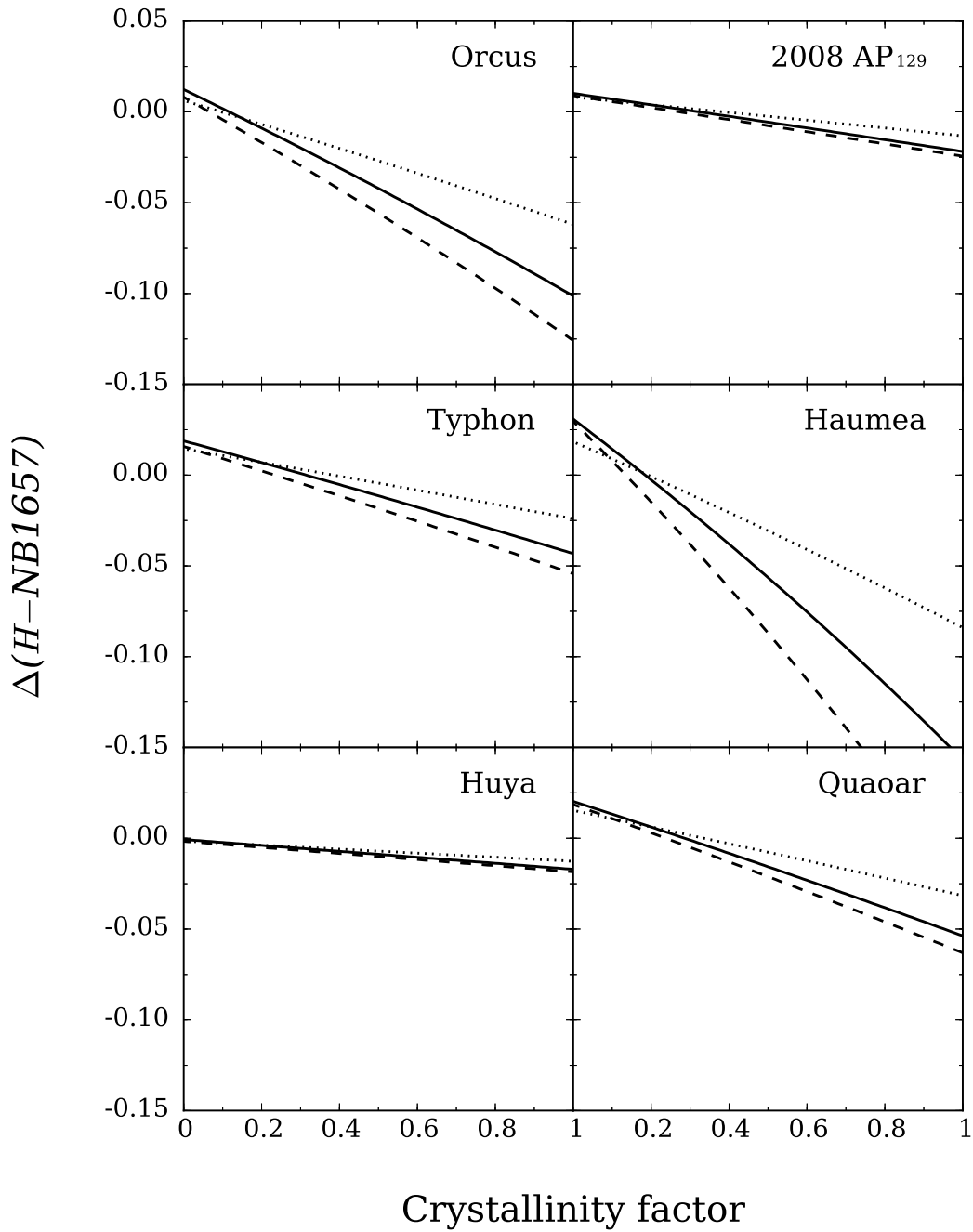}
\caption{
 The relations between $H$$-$NB1657 index and crystallinity factor estimated
 from the model spectra with grain size of 10~$\micron$ (dotted lines), 
 50~$\micron$ (solid lines), and 200~$\micron$ (dashed lines).
 \label{fig05}}
\end{figure}

\begin{figure}
\epsscale{1.00}
\plotone{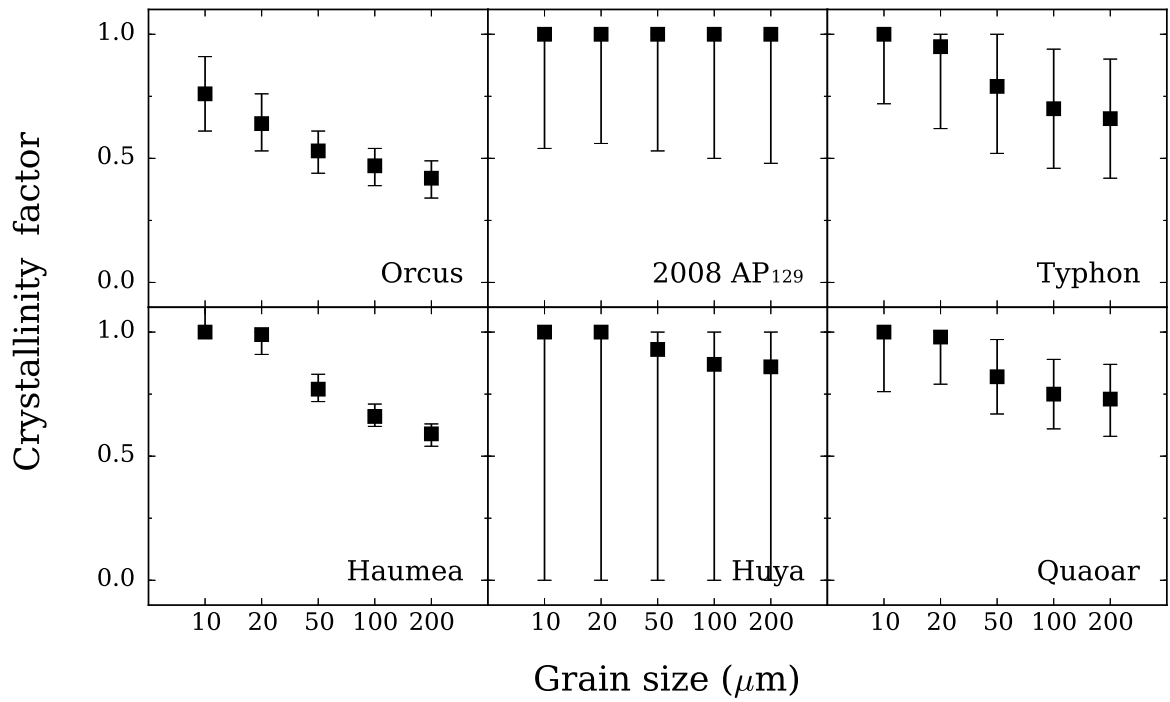}
\caption{
 Dependency of the obtained crystallinity factors on assumed grain size from 10~$\micron$ to
 200~$\micron$ in diameter.
 The error bars show the 1-$\sigma$ uncertainty.
 \label{fig06}}
\end{figure}

We examined the variation of crystallinity factor for our observed objects among the spectral
models with grain sizes from 10~$\micron$ to 200~$\micron$ in diameter (see Figure~\ref{fig05}).
A uniform size between the crystalline and amorphous particles was assumed in each pattern.
The results are shown in Figure~\ref{fig06}.
One can see that the crystallinity factor monotonically decreases with increasing grain size.
This is because of the optical characteristic of H$_2$O ice that larger grains increase the amount
of 1.65-$\micron$ absorption rather than that of entire absorption covered by the $H$ band, as
pointed out in \cite{1998JGR...10325809G}.
Although the crystallinity factor varies greatly between the smallest and largest grain sizes,
it remains larger than $\sim$0.5 within the error in Haumea, Quaoar, Typhon, and
2008~AP$_{129}$.
Orcus has a slightly lower crystallinity in case of the largest grain sizes, but definitely
contains a certain level of crystalline ice.
This result indicates that all of the five objects whose crystallinity has been precisely
determined are likely to be covered by a surface containing crystalline-dominant H$_2$O ice or
comparable to it.

\subsection{Impurities}\label{sec:subsec3-4}

The spectrum model given by Equation~(\ref{eq2}) approximates the surface reflectance spectrum of
the target objects as a combination of H$_2$O ice and other materials of which the total spectrum
shows a linear-sloped continuum with no absorption feature.
For the $H$-band data, this assumption seems to be reasonable because most non-H$_2$O ices which
are potentially contained on the TNOs' surface, e.g., CH$_3$OH, CO, CO$_2$, N$_2$, and NH$_3$, have no
or only slight absorptions over the wavelength range
\citep{2005ApJ...620.1140G,1984Icar...58..293C}.
However, CH$_4$ ice could be influential for determination of H$_2$O ice crystallinity because
it exhibits several absorption features from 1.5~$\micron$ to
1.8~$\micron$ \citep{1991JGR....96..477P}.
In particular, the strong absorption bands at 1.67~$\micron$ possibly makes a significant negative
contribution to the $\Delta$($H$$-$NB1657) index.

Various near-infrared spectroscopies suggest the presence of CH$_4$ ice on several TNOs with
H$_2$O ice-rich surface including Quaoar \citep{2007ApJ...670L..49S,2009A&A...501..349D}
and Orcus \citep{2008A&A...479L..13B,2010A&A...520A..40D,2011A&A...534A.115C}.
According to previous studies of the two spectra, the average fraction of CH$_4$ ice is $\sim$0.05,
corresponding to CH$_4$/H$_2$O~$\sim$~0.15.
To evaluate the effect of CH$_4$ ice mixing on measurements of H$_2$O ice crystallinity, we
calculated the $\Delta$($H$$-$NB1657) index with several patterns of the particle number ratio
(CH$_4$/H$_2$O~=~0.05--0.20) and CH$_4$ ice grain size ($d$~= 10, 50, and 100~$\micron$).
The optical constants of CH$_4$ ice used for the spectrum modeling are as follows: 
(i) the real part of refractive index $n$ is given by $n = 1.38 - 0.03\lambda$
\citep[0.67~$\micron$~$\leq$~$\lambda$~$\leq$~2.0~$\micron$; ][]{2008ssbn.book..483D},
(i\hspace{-1pt}i) the imaginary part of refractive index $k$ is converted from the absorption
coefficients of phase~I CH$_4$ ice (stable between 20~K and 90~K) presented by
\cite{2002Icar..155..486G}.
The synthetic model spectra were generated from Equation~(\ref{eq2}) by replacing the reflectance
spectrum of H$_2$O ice, $S_{\rm H_2O}$, with that of mixed ices of H$_2$O and CH$_4$ derived from
the intimate mixture model developed by \cite{1993tres.book.....H}.
The grain size of H$_2$O ice was fixed to $d$~=~50~$\micron$.

\begin{figure}
\epsscale{0.90}
\plotone{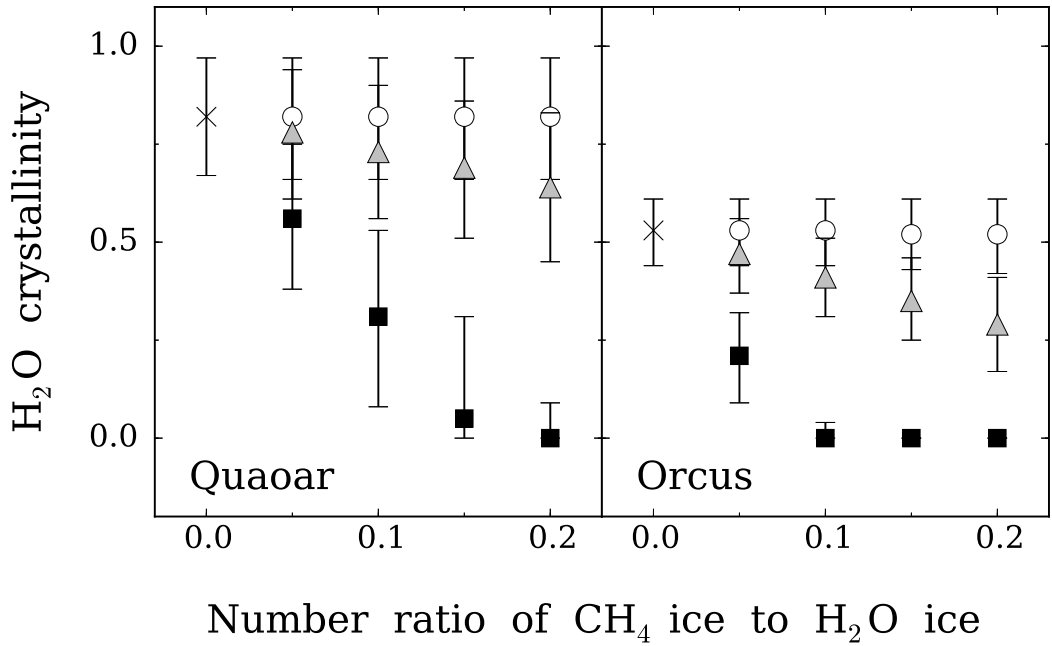}
\caption{
 Dependency of the obtained crystallinity factors on the particle number ratio of CH$_4$/H$_2$O.
 The grain sizes of CH$_4$ ice are 10~$\micron$ (open circles), 50~$\micron$ (triangles), and
 100~$\micron$ (squares) in diameter.
 The grain size of H$_2$O ice is assumed to be 50~$\micron$ in diameter.
 The crosses show the case of absence of CH$_4$ ice.
 The error bars show the 1-$\sigma$ uncertainty.
 \label{fig07}}
\end{figure}

The results are shown in Figure~\ref{fig07}.
If CH$_4$ ice grain is equal to or smaller than H$_2$O ice grain, the H$_2$O ice crystallinity
factors are larger than 0.50 and 0.25 for Quaoar and Orcus, respectively.
The situation is similar if CH$_4$/H$_2$O~$\lesssim$~0.05.
In contrast, the absorption due to the larger-grain CH$_4$ ice on the surface with
CH$_4$/H$_2$O~$\gtrsim$~0.1 dominates at $\sim$1.65~$\micron$ and could cause a great
uncertainty of H$_2$O ice crystallinity.

We recognize that \cite{2009A&A...501..349D} and \cite{2011A&A...534A.115C} provide the most
reliable spectral modeling for Quaoar and Orcus, respectively, among the published works.
The former suggests Quaoar's surface containing fine CH$_4$ ice grains of $d$~$\sim$~10~$\micron$
with a slightly high fraction of CH$_4$/H$_2$O~$\sim$~0.3.
The latter suggests Orcus's surface containing coarse CH$_4$ ice grains of
$d$~$\sim$~100~$\micron$ with a tiny fraction of CH$_4$/H$_2$O~$<$~0.03.
If those parameters are simply applied to our modeling, they indicate that the potential presence
of CH$_4$ ice is unlikely to induce significant overestimation of H$_2$O ice crystallinity against
Quaoar and Orcus.

\section{DISCUSSION}\label{sec:section}

Provided that the observed objects are coated by H$_2$O ice grain of $d$~$\lesssim$~50~$\micron$
with negligible contamination by other ices such as CH$_4$, our results suggest the following
implications:

\noindent (i) Not only Haumea and Quaoar known as large objects rich in crystalline H$_2$O ice, but
also smaller objects, Typhon and 2008~AP$_{129}$, contain crystalline-dominated H$_2$O ice on the
surfaces.

\noindent (i\hspace{-0.5pt}i) Orcus has a crystallinity of comparable to 0.5, obviously lower than
Haumea and Quaoar.

\noindent (i\hspace{-1.0pt}i\hspace{-1.0pt}i) No objects with low-crystallinity surface have been
found in our well-measured TNO sample.

The initial state of H$_2$O ice in TNOs is likely to be amorphous if the bodies formed beyond
$\sim$25~au from the Sun \citep{2006AdSpR..38.1968K}.
Such pristine ice could be heated and crystallized through thermal evolution after the formation.
Based on the timescale to onset of crystallization \citep{1996ApJ...473.1104J} and the timescale to
completion of crystallization \citep{1994A&A...290.1009K}, amorphous ice at $>$~70~K would be
crystallized within the solar system age \citep{2013sssi.book..371M}.
Additionally, surface annealing due to micrometeorite impacts may also be effective in inducing
ice crystallization.
In this section, we discuss several possible mechanisms for crystallization of the surface H$_2$O
ice of TNOs.

Note that crystalline H$_2$O ice can be amorphized by irradiation of UV photon as well as
energetic particles such as protons and electrons from the solar wind and cosmic rays.
\cite{2003A&A...397....7L} and \cite{2005MSAIS...6...57L} confirmed that UV photolysis induces
the amorphization at 16 and 90~K, respectively, through irradiation experiments.
However, as \cite{2008ssbn.book..507H} pointed out, the UV penetration depth is much less than
the optical depth in near-infrared ($\sim$350~$\micron$).
The effect of UV photolysis can be excluded in this observation.
On the other hand, protons above $\sim$1~MeV and electrons above $\sim$0.1~MeV have stopping
ranges larger than the optical depth for H$_2$O ice target \citep{2008ssbn.book..507H}.
\cite{2006Icar..183..207M} presented the results of proton irradiation experiments, indicating
(i) amorphous ice was produced at low temperatures ($<$~40~K),
(i\hspace{-0.5pt}i) some crystalline ice persisted at 50~K after irradiation, and
(i\hspace{-1.0pt}i\hspace{-1.0pt}i) the crystalline spectrum showed only slight changes at
$\geq$~70~K.

\cite{2009JPCA..11311174Z} investigated the effect of electron irradiation on the near-infrared
spectra and found that crystalline H$_2$O ice can be converted only partially to amorphous at
40~K or higher.
According to measurements from the Geostationary Operational Environmental
Satellites (GOES)\footnote{The data distributed by NOAA/National Centers for Environmental
Information (\url{ftp.ngdc.noaa.gov}).}, the fluence ratio of electrons to protons above 1~MeV
from solar wind is 10$^3$--10$^4$.
The interplanetary flux of proton from cosmic ray is $\sim$10$^2$ greater than that from solar
wind above 1~MeV at heliocentric distance of 40~au \citep{2003EM&P...92..261C}.
Thus, the electron is likely to be the dominant component of energetic particle irradiance on
surfaces of icy bodies in the main trans-neptunian region.
Although there are a lot of discrepancies among previous experimental studies as mentioned in
\cite{2013sssi.book..371M}, assuming the irradiation tolerance of crystalline H$_2$O ice 
based on the experimental results of \cite{2009JPCA..11311174Z}, the amorphization process could
be only limited.
Here we ignore the effect of radiation-induced amorphization in the following discussion.

\subsection{Insolation}\label{sec:subsec4-1}

Prior to discussions of the thermal evolution, we should confirm the variations in surface
temperature of the target objects over an orbit.
Amorphous ice would be crystallized on surfaces that have been much warmer than 70~K
over the timescale of the solar system age or less
\citep{1996ApJ...473.1104J,1994A&A...290.1009K}.
Even if there is no heat source except solar insolation, the transition from amorphous to
crystalline occurs on the surface of bodies sufficiently close to the Sun.
All the target objects other than Typhon have perihelion distances larger than 28.5~au,
which are converted into surface temperature less than 66~K by Equation~(\ref{eq4}).
Assuming no experiences of significant migration from the formation, their surfaces have been cold
enough for amorphous ice to be continuously stable.

However, Typhon has a high orbital eccentricity (0.54) allowing it to approach up to 17.5~au to the
Sun.
Around its perihelion, the sunward surface would be warmed up to $\sim$86~K, while the ordinary
temperature is $\sim$50~K during most of the orbit.
In the model of \cite{1996ApJ...473.1104J}, the time to onset of crystallization at the temperature
is $\sim$10$^6$~yr.
The insolation heating can induce surface crystallization if Typhon keeps such an eccentric orbit
over the age of the solar system and the intermittent contribution has monotonically accumulated.

\subsection{Radiogenic Heating}\label{sec:subsec4-2}

The decay of radioactive isotopes contained in dust mixed with ice is the most important heat
source for the internal thermal evolution of small icy bodies in the outer solar system.
It allows the crystallization to proceed outward from the deep interior of the body.

In \cite{1995Icar..117..420P}, thermal evolutionary calculations of icy bodies larger than 40~km in
diameter ($D$) with heat sources of $^{26}$Al and four long-lived radioactive isotopes indicate
that H$_2$O ice on the outer layer seems to be left in the amorphous form.
\cite{2002Icar..160..300C} also reported that $^{26}$Al decay could not raise the surface
temperature of TNOs with $D$~=~20--1000~km higher than $\sim$50~K and not induce H$_2$O ice
crystallization at the surface layers.
Those models cannot explain surface crystallization of objects with $D$~$>$~100~km.

However, assuming that the objects were formed closer to the Sun than the present orbits,
radiogenic heating can potentially crystallize the surface ice because of fast accretional growth,
thus sufficient radiogenic heating by $^{26}$Al.
The simulations of \cite{2006Icar..183..283M} showed that icy bodies above $D$~$\sim$~30~km
originally located around 20~au could be covered by a crystalline ice surface.
\cite{2008ssbn.book..213M} showed that if icy bodies formed at heliocentric distance less than
$\sim$32~au, the crystalline/amorphous ice boundary reaches the surface.

More recent work by \cite{2011A&A...529A..71G} presented three-dimensional simulations to compute
the thermal evolution of icy small bodies with heat sources of several short- and long-lived
radioactive isotopes.
Their results suggest:

\begin{itemize}
 \item  When the growth advanced quickly ($\sim$1~Myr) and short-lived isotope decay
        effectively contributes to the thermal evolution, objects with $D$~$>$~100~km retain only a
        thin ($<$~2~km) layer of amorphous H$_2$O ice on their surface, that may possibly be
        removed by impact excavating.
        In addition, provided such objects are formed close enough to the Sun (10--15~au),
        crystallization would be triggered at the surface by insolation.
 \item  When the thermal evolution is dominated by long-lived isotope decay, only large objects
        ($D$~$>$~600~km) with a bulk density of at least 1500~kg~m$^{-3}$ should contain
        crystalline H$_2$O ice close to the surface.
\end{itemize}

Our findings regarding the presence of crystalline H$_2$O ice on Typhon and 2008~AP$_{129}$ agree
with the first scenario.
In contrast, Typhon and possibly 2008~AP$_{129}$ (see Table~\ref{tab1}) are inconsistent with the
second one.
Based on the model of \cite{2011A&A...529A..71G}, at least those small objects should have grown in
a sufficiently short time scale for the surfaces to be heated up to the crystallization temperature
by the decay of short-lived isotopes unless they were formed close to the Sun (10--15~au).

Such a rapid formation as to induce a sufficient heating for surface ice crystallization due to the
decay of short-lived isotopes also can occur at a smaller heliocentric distance than the current
location.
Actually, several theoretical studies support this model.
\cite{2004come.book...97W} presented coagulation simulations of planetesimal growth in the outer
solar system and showed the formation of objects with $D$~$>$~100~km only inside $\sim$30~au.
More recently, \cite{2012AJ....143...63K} performed $N$-body coagulation calculations for the
formation and evolution of TNOs, indicating that objects at 15--18~au grow up to 100~km in diameter
within $\sim$1~Myr at earliest, while it takes more than 10~Myr at 33--40~au.
Outward transport of planetesimals from the inner regions via the planetary migration has been
believed to take place based on numerical simulations for dynamical evolutions of giant planets and
TNOs
\citep[e.g., ][]{2003Icar..161..404G,2003Natur.426..419L,2005AJ....130.2392H,2008Icar..196..258L}.
The rapid growth around 20~au from the Sun is a plausible mechanism for crystallization of
surface ice from the early radioactive heating on TNOs as small as Typhon.

The thermal evolution model may also be able to explain the low fraction of crystalline ice on
Orcus compared with Haumea and Quaoar.
One of the important parameters that the heating rate depends on is bulk density of the body.
A higher density increases the abundance of dust materials containing radioactive isotopes and thus
enhances heating efficiency.
The difference in bulk density possibly causes the variation in crystallinity among comparable-size
objects.
\cite{2011A&A...529A..71G} suggest that a density of at least 1.5~g~cm$^{-3}$ is required for the
presence of crystalline ice on the surface of objects with $D$~$\gtrsim$~600~km.
Haumea is well known to have a high density
\citep[$\sim$~2.6~g~cm$^3$;][]{2006ApJ...639.1238R, 2007AJ....133.1393L}.
Quaoar's density also has been estimated to be high as 
2.18$^{+0.43}_{-0.36}$~g~cm$^{-3}$ from thermal flux \citep{2013A&A...555A..15F}
or 1.99~$\pm$~0.46~g~cm$^{-3}$ from occultation \citep{2013ApJ...773...26B}.
In contrast, Orcus exhibits an obviously lower density of $\sim$1.5~g~cm$^{-3}$
\citep{2010AJ....139.2700B}, just about the same value as the criteria for
surface ice crystallization shown by \cite{2011A&A...529A..71G}.
This fact suggests the possibility that Orcus failed to take a sufficient rise in temperature
of the upper layer for crystallizing most of the surface H$_2$O ice due to a deficiency in
the amount of radioactive isotopes.
The potential correlation between crystallinity and bulk density could support the hypothesis that
radiogenic heating is the dominant effect on surface crystallization for TNOs.

\subsection{Cryovolcanism}\label{sec:subsec4-3}

A portion of surface ice may not have been formed in-situ but been supplied through the interior by
geological activity of ice volcanic eruption, so-called cryovolcanism.
Actually, atmospheric plumes have been shown on several icy satellites of giant planets,
such as Triton \citep{1989Sci...246.1422S, 1994EM&P...67..101K},
Enceladus \citep{2006Sci...311.1393P, 2009sfch.book..683S}, and 
Europa \citep{2014Sci...343..171R}, as possible evidence of cryovolcanic events.
The occurrence of substantial liquid H$_2$O flow to the surface requires some sort of continual
heating process to melt the subsurface ice.
For the satellites, tidal energy dissipation is likely to be a major heat source
\citep[e.g.,][]{2000JGR...10529283R}.
Even though inclusion of NH$_3$ in H$_2$O ice allows the melting point to drop from 273~K to 176~K
at minimum \citep{1991Icar...89...93K} and decreases the thermal conductivity, it is still
uncertain whether the cryovolcanic activity can be driven on TNOs primarily by long-lived
radioactive isotopes without tidal heating.

\cite{2007ApJ...663.1406C} suggested that a body with $D$~$>$~1200~km composed of rock with a mass
fraction $\gtrsim$~0.7 (corresponding to a bulk density $\gtrsim$~1.8~g~cm$^{-3}$) and NH$_3$-rich
H$_2$O ice ($\sim$0.15 by weight) could develop cryovolcanism and be steadily resurfaced by
crystalline H$_2$O ice with a sufficient rate to be detected by near-infrared wavelength.
Haumea, with a mean diameter of 1200--1300~km \citep{2010A&A...518L.147L,2013A&A...555A..15F}
and a density of 2.6~g~cm$^{-3}$ \citep{2014EM&P..111..127L}, satisfies those conditions.
Quaoar is slightly smaller than the criteria size ($\sim$1000--1100~km) and has a high density
(1.99~$\pm$~0.46~g~cm$^{-3}$; \citealp{2013ApJ...773...26B}, 2.18$^{+0.43}_{-0.36}$~g~cm$^{-3}$;
\citealp{2013A&A...555A..15F}).
Orcus is likely to be comparable in size to Quaoar
\citep[$\sim$850--950~km;][]{2010AJ....139.2700B,2010A&A...518L.148L,2013A&A...555A..15F},
while the density is relatively low
\citep[1.53$^{+0.15}_{-0.13}$~g~cm$^{-3}$;][]{2013A&A...555A..15F}.
As pointed out by \cite{2009Icar..202..694D}, the ice containing CH$_3$OH along with NH$_3$
freezes at lower temperature \citep[153~K;][]{1994EM&P...67..101K} and therefore is likely to
reduce the minimum diameter of a TNO which can retain liquid H$_2$O to 800--1000~km.
Cryovolcanism may be active on Quaoar and Orcus if they consist of CH$_3$OH-NH$_3$-H$_2$O ice.

Unfortunately, our results cannot provide explicit constraints on the source of crystalline
H$_2$O ice at the surfaces of Haumea, Quaoar, and Orcus.
On the other hand, Typhon and 2008~AP$_{129}$ are likely to be too small to exhibit cryovolcanic
activity even if antifreeze compounds such as NH$_3$ and CH$_3$OH are sufficiently mixed in the
subsurface ice.

\subsection{Other mechanisms}\label{sec:subsec4-4}

Instead of the internal heat sources as mentioned above, H$_2$O ice crystallization on surfaces of
airless bodies in the trans-neptunian region could be induced by impacts of meteorites such as
interplanetary dust particles (IDPs).
\cite{2007ApJ...663.1406C} assessed possible mechanisms with micrometeorites for explaining the
presence of crystalline H$_2$O ice on Charon's surface, namely, impact gardening and annealing.
They approximated the mass flux of IDPs from Pioneer~10 observations as
2.4$~\times$~10$^{-17}$~kg~s$^{-1}$~m$^{-2}$ at 18~au and concluded that those mechanisms
make little contribution to surface renewal on either Charon or other TNOs.
In contrast, \cite{2010Icar..208..492P} expected the IDP flux at Uranus to be 
1.2$~\times$~10$^{-16}$~kg~s$^{-1}$~m$^{-2}$ and claimed that the impact annealing could
be effective for surface crystallization on TNOs.

As seen in Table~2 of \cite{2010Icar..208..492P}, the impact velocity of IDPs and required dust
flux for annealing are similar among objects in the main trans-neptunian belt.
Assuming homogeneous IDP distribution over the region, surface crystallinity would be almost
uniform among TNOs if micrometeorite impacts are the primary factor.
However, considering our result indicating a low crystallinity of Orcus compared with Haumea and
Quaoar, micrometeorite annealing seems not to be a major mechanism of surface crystallization at
least for icy objects as large as Orcus.

Finally, we focus on the unique situation of 2008~AP$_{129}$ that has an orbit consistent with the
Haumea collisional family \citep{2007Natur.446..294B}.
The relative velocity to the estimated collision's center of mass, $\sim$140~m~s$^{-1}$, also
agrees with the velocity dispersion of the known family members (50--300~m~s$^{-1}$), suggesting
a dynamical association with the Haumea family \citep{2012Icar..221..106V}.
The spectra of family members uniquely show significantly high fractions of H$_2$O ice as well as
the crystalline feature \citep{2008AJ....135...55B, 2011Icar..214..297B, 2012AJ....143..146B}.
However, 2008~AP$_{129}$ contains only a minor fraction of H$_2$O ice
\citep[$f_{\rm H_2O}$ = 0.14\,$\pm$\,0.09;][]{2012AJ....143..146B},
which prevents characterization of the membership.
If 2008~AP$_{129}$ originates from a collisional fragment as an ice-poor family member, the
presence of crystalline H$_2$O ice on the surface is natural because the body is likely to derive
from the icy mantle of Haumea.
The lack of ice could be explained from inhomogeneous compositions of the partial differentiation
\citep{2012Icar..221..106V} although the actuality is still unknown.

\section{CONCLUSIONS}\label{sec:conclusions}

We develop a new technique for measuring the crystallinity of H$_2$O ice on the surface of TNOs
with near-infrared narrow-band photometry using Subaru/MOIRCS.
The strength of the 1.65-$\micron$ absorption band for five objects has been obtained and converted
into the fraction of crystalline ice by comparing with the spectrum model.
The largest objects, Haumea, Quaoar, and Orcus, show a crystalline-rich icy surface as many
previous spectroscopic studies have reported, while it is also found out that their surfaces are
likely to contain amorphous ice unless the grain size is smaller than $\sim$20~$\micron$.
For Haumea and Quaoar, the model spectra based on the determined fractions of crystalline ice
well agree with the published near-infrared spectra, indicating reasonable accuracy of our
measurements.

The results indicate that H$_2$O ice on Haumea and Quaoar are highly dominated by the crystalline
state, while Orcus shows a higher fraction of amorphous ice.
Based on the model of thermal evolution due to radiative decay, the low bulk density of Orcus
could cause suppression of the surface heating and stagnation in crystallization.
It is of great significance to examine the possibility of the positive correlation between bulk
density of the body, i.e., abundance of refractory inclusion, and surface crystallinity,
providing a critical clue for understanding the crystallization mechanism.

We also found the presence of crystalline H$_2$O ice on Typhon and 2008~AP$_{129}$.
Those objects are smaller than the expected critical size ($D$~$\sim$~600~km) for surface
crystallization assuming that they were formed far enough from the Sun ($>$~15~au) and the initial
thermal evolution was dominated by the decay of long-lived isotopes \cite{2011A&A...529A..71G}.
It is still uncertain which thermal or non-thermal process is the primary crystallization
mechanism for midsize TNOs.
Further investigations are required to determine whether the body size affects the production of
crystalline H$_2$O ice at the surface layer.

\acknowledgments

We are grateful to Tadayuki Kodama who kindly provided the NB1657 filter to open-use observers.
We thank Ichi Tanaka for technically supporting our observation and data reduction, as well as
Naruhisa Takato for fruitful comments on the manuscript.
We also thank an anonymous referee for helpful comments and suggestions.
This study is based on data collected at Subaru Telescope, National Astronomical Observatory of
Japan.


\clearpage


\clearpage

\begingroup
\renewcommand{\arraystretch}{1.2}
\begin{table}
\caption{
 Orbital elements (semi-major axis, $a$, eccentricity, $e$, and inclination, $i$), absolute visual
 magnitude ($\mathcal{H}_V$), diameter ($D$), and spectral parameters (amount of H$_2$O ice, 
 $f_{\rm H_2O}$
 and slope of the continuum, $m_{\rm cont}$; Brown et al. 2012) of observed objects.
 \label{tab1}}
\begin{tabular}{lccccccc}
\\
\hline\hline
\hspace{2em} Object       & $a$  & $e$ & $i$   & $\mathcal{H}_V$ & $D$  & $f_{\rm H_2O}$ & $m_{\rm cont}$ \\
                          & (au) &     & (deg) & (mag)           & (km) &               &                \\
\hline
\ (38628) Huya            & \, 39.41 & 0.276 &   15.5 & 4.9
                          & \ 458\,$\pm$\,\,9\ $\tablenotemark{a}$
                          & 0.08\,$\pm$\,0.02 &  $-$0.09\,$\pm$\,0.00 \\
\ (42355) Typhon          & \, 38.11 & 0.540 & \, 2.4 & 7.5
                          & \ 185\,$\pm$\,\,7\ $\tablenotemark{b}$
                          & 0.31\,$\pm$\,0.17 & \ \ 0.05\,$\pm$\,0.10 \\
\ (50000) Quaoar          & \, 43.22 & 0.035 & \, 8.0 & 2.4
                          & 1074\,$\pm$\,38$\tablenotemark{a}$
                          & 0.29\,$\pm$\,0.01 & \ \ 0.07\,$\pm$\,0.00 \\
\ (65489) Ceto            & 101.91 & 0.825 &   22.3 & 6.4
                          & \ 281\,$\pm$\,11$\tablenotemark{b}$
                          & 0.17\,$\pm$\,0.28 &  $-$0.10\,$\pm$\,0.17 \\
\ (90482) Orcus           & \, 39.44 & 0.219 &   20.5 & 2.2
                          & \ 958\,$\pm$\,23$\tablenotemark{a}$
                          & 0.44\,$\pm$\,0.01 &  $-$0.36\,$\pm$\,0.03 \\
 (136108) Haumea          & \, 43.17 & 0.192 &   28.2 & 0.1
                          & 1240 $^{+69}_{-58}$\,$\tablenotemark{a}$
                          & 0.66\,$\pm$\,0.00 &  $-$0.40\,$\pm$\,0.00 \\
 (315530) 2008~AP$_{129}$ & \, 42.07 & 0.143 &   27.4 & 4.9
                          & \ \ 471 $^{+161}_{-79}$\,$\tablenotemark{c}$
                          & 0.14\,$\pm$\,0.09 & \ \ 0.04\,$\pm$\,0.05 \\
\hline
\end{tabular}
\tablenotetext{a}{\cite{2013A&A...555A..15F}.}
\tablenotetext{b}{\cite{2012A&A...541A..92S}.}
\tablenotetext{c}{Estimated from $\mathcal{H}_V$ assuming a geometric albedo of 0.09$\pm$0.04, the
average value of TNOs from \cite{2012A&A...541A..92S}.}
\end{table}
\endgroup

\clearpage

\begingroup
\renewcommand{\arraystretch}{1.2}
\begin{table}
\caption{ 
 Observational circumstances.
 \label{tab2}}
\begin{tabular}{lccccc}
\\
\hline\hline
\hspace{2em} Object       & UT & Airmass & Seeing$\tablenotemark{a}$ & $H$$\tablenotemark{b}$
                          & NB1657$\tablenotemark{c}$ \\
                          &    &         & (arcsec)                  & (sec)                 
                          & (sec)                       \\
\hline
\ (90482) Orcus           & 07:28--08:28 & 1.13--1.19 & 0.64 & 880 &   1260 \\
 (315530) 2008~AP$_{129}$ & 08:34--09:38 & 1.20--1.38 & 0.64 & 840 &   1440 \\
\ (42355) Typhon          & 10:08--11:12 & 1.10--1.17 & 0.75 & 780 &   1440 \\
\ (65489) Ceto            & 11:53--12:29 & 1.34--1.46 & 0.64 & 500 & \, 720 \\
 (136108) Haumea          & 12:34--13:24 & 1.06--1.16 & 0.66 & 600 & \, 810 \\
\ (38628) Huya            & 13:50--14:53 & 1.14--1.27 & 0.76 & 840 &   1440 \\
\ (50000) Quaoar          & 14:57--15:42 & 1.22--1.25 & 0.78 & 450 &   1080 \\
\hline
\end{tabular}
\tablenotetext{a}{Typical full width at half maximum of point sources.}
\tablenotetext{b}{Total exposure time in the $H$ filter.}
\tablenotetext{c}{Total exposure time in the NB1657 filter.}
\end{table}
\endgroup

\clearpage

\begingroup
\renewcommand{\arraystretch}{1.2}
\begin{table}
\caption{
 Results of Photometry.
 \label{tab3}}
\begin{tabular}{lcc}
\\
\hline\hline
\hspace{2em}Object        & $H$$\tablenotemark{a}$ & $\Delta$($H$--NB1657)$\tablenotemark{a}$ \\
                          & (mag) & (mag)                 \\
\hline
\ (90482) Orcus           & 17.89 $\pm$ 0.03  & -0.045 $\pm$ 0.010 \\
 (315530) 2008~AP$_{129}$ & 19.19 $\pm$ 0.03  & -0.031 $\pm$ 0.024 \\
\ (42355) Typhon          & 18.24 $\pm$ 0.03  & -0.030 $\pm$ 0.017 \\
\ (65489) Ceto            & 18.85 $\pm$ 0.03  & +0.033 $\pm$ 0.032 \\
 (136108) Haumea          & 16.37 $\pm$ 0.03  & -0.109 $\pm$ 0.011 \\
\ (38628) Huya            & 17.57 $\pm$ 0.03  & -0.016 $\pm$ 0.016 \\
\ (50000) Quaoar          & 16.74 $\pm$ 0.03  & -0.040 $\pm$ 0.012 \\
\hline
\end{tabular}
\tablenotetext{a}{The errors show the 1-$\sigma$ uncertainty.}
\end{table}
\endgroup

\clearpage

\clearpage

\begingroup
\renewcommand{\arraystretch}{1.5}
\begin{table}
\caption{
 Estimated crystallinity factors with grain size of 50~$\micron$ in diameter.
 \label{tab4}}
\begin{tabular}{lcclc}
\\
\hline\hline
\hspace{2em} Object & $r_{\rm h}$$\tablenotemark{a}$ & $T$$\tablenotemark{b}$ &
\vspace{-0.7em}
\hspace{2em} H$_2$O ice spectra$\tablenotemark{c}$ & Crystallinity factor$\tablenotemark{d}$\\
                          & (au) & (K) &                           &                        \\
\hline
\ (90482) Orcus           & 48.0 &  49 & \ crys\_50K, amorph\_low  & 0.53$^{+0.08}_{-0.09}$ \\
 (315530) 2008~AP$_{129}$ & 37.9 &  56 & \ crys\_60K, amorph\_low  & 1.00$^{+0.00}_{-0.47}$ \\
\ (42355) Typhon          & 19.1 &  82 & \ crys\_80K, amorph\_high & 0.79$^{+0.21}_{-0.27}$ \\
 (136108) Haumea          & 50.8 &  48 & \ crys\_50K, amorph\_low  & 0.77$^{+0.06}_{-0.05}$ \\
\ (38628) Huya            & 28.6 &  66 & \ crys\_70K, amorph\_low  & 0.93$^{+0.07}_{-0.93}$ \\
\ (50000) Quaoar          & 43.1 &  52 & \ crys\_50K, amorph\_low  & 0.82$^{+0.15}_{-0.15}$ \\
\hline
\end{tabular}
\tablenotetext{a}{Heliocentric distance at the observation.}
\tablenotetext{b}{Thermal temperature from Equation~(\ref{eq4}).}
\tablenotetext{c}{The optical constant dataset derived from laboratory spectra provided
by \cite{2008Icar..197..307M}.}
\tablenotetext{d}{The errors show the 1-$\sigma$ uncertainty.}
\end{table}
\endgroup

\clearpage

\clearpage

\begingroup
\renewcommand{\arraystretch}{1.5}
\begin{table}
\caption{
 Dependency of estimated H$_2$O ice fraction $f_{\rm H_2O}$ on crystallinity $f_{\rm crys}$
 (see text).
 \label{tab5}}
\begin{tabular}{lcccccc}
\\
\hline\hline
          & $f_{\rm crys}$ & & & \multicolumn{2}{c}{$f_{\rm H_2O}$} & \\
\cline{2-2} \cline{4-7}
 Given    & 1.00 & & 0.70 & 0.50 & 0.30 & 0.10 \\
\hline
          & 0.50 & & 0.69 & 0.50 & 0.30 & 0.10 \\
 Best-fit & 0.25 & & 0.69 & 0.49 & 0.29 & 0.10 \\
          & 0.00 & & 0.68 & 0.48 & 0.29 & 0.10 \\
\hline
\end{tabular}
\end{table}
\endgroup

\clearpage

\begingroup
\renewcommand{\arraystretch}{1.5}
\begin{table}
\caption{
 Estimated crystallinity factors with lower temperatures.
 \label{tab6}}
\begin{tabular}{lclc}
\\
\hline\hline
\hspace{2em} Object & $T$$\tablenotemark{a}$ &
\vspace{-0.7em}
\hspace{2em} H$_2$O ice spectra$\tablenotemark{b}$ & Crystallinity factor$\tablenotemark{c}$\\
                          & (K) &                           &                        \\
\hline
\ (90482) Orcus           &  42 & \ crys\_40K, amorph\_low  & 0.51$^{+0.09}_{-0.08}$ \\
 (315530) 2008~AP$_{129}$ &  48 & \ crys\_50K, amorph\_low  & 1.00$^{+0.00}_{-0.47}$ \\
\ (42355) Typhon          &  70 & \ crys\_70K, amorph\_high & 0.72$^{+0.24}_{-0.24}$ \\
 (136108) Haumea          &  41 & \ crys\_40K, amorph\_low  & 0.75$^{+0.06}_{-0.05}$ \\
\ (38628) Huya            &  56 & \ crys\_60K, amorph\_low  & 0.89$^{+0.11}_{-0.89}$ \\
\ (50000) Quaoar          &  44 & \ crys\_40K, amorph\_low  & 0.80$^{+0.15}_{-0.15}$ \\
\hline
\end{tabular}
\tablenotetext{a}{Reduced thermal temperature (see text).}
\tablenotetext{b}{The optical constant dataset derived from laboratory spectra provided
by \cite{2008Icar..197..307M}.}
\tablenotetext{c}{The errors show the 1-$\sigma$ uncertainty.}
\end{table}
\endgroup

\clearpage

\clearpage

\clearpage

\clearpage

\end{document}